\documentclass[final,1p,times]{elsarticle}

\usepackage{amssymb}
\usepackage{mathtools, cuted}
\usepackage{epsfig}
\usepackage[english]{babel}
\usepackage{physics}
\usepackage{color}
\usepackage{subcaption}
\usepackage{hyperref}
\usepackage{lipsum}
\usepackage{cases}
\usepackage{bbold}

\newcommand*{\lop}{\mathcal{L}_{t,x}\,}
\newcommand*{\lopt}{\mathcal{L}_{t}\,}
\newcommand*{\nlop}{\mathcal{N}_{t,x}\,}
\newcommand*{\nloptilde}{\widetilde{\mathcal{N}}^{[n]}_{t,x}\,}
\newcommand*{\rop}{\mathcal{R}_{x}\,}
\newcommand*{\ropx}{\mathcal{R}_{x'}\,}
\newcommand{\vect}[1]{\boldsymbol{#1}}
\newcommand{\me}{\mathrm{e}}
\usepackage{amsthm}

\journal{Partial Differential Equations in Applied Mathematics}

\begin{document}

\begin{frontmatter}

\title{The BLUES function method for second-order partial differential equations: application to a nonlinear telegrapher equation}

\author[1]{Jonas Berx\corref{cor1}}
\ead{jonas.berx@kuleuven.be}
\author[1]{Joseph O. Indekeu}
\ead{joseph.indekeu@kuleuven.be}
\cortext[cor1]{Corresponding author}
\address[1]{Institute for  Theoretical Physics, KU Leuven, B-3001 Leuven, Belgium}

\date{\today}

\begin{abstract}
An analytic iteration sequence based on the extension of the BLUES (Beyond Linear Use of Equation Superposition) function method to partial differential equations (PDEs) with second-order time derivatives is studied. The original formulation of the BLUES method is modified by introducing a matrix formalism that takes into account the initial conditions for higher-order time derivatives. The initial conditions of both the solution and its derivatives now play the role of a source vector. The method is tested on a nonlinear telegrapher equation, which can be reduced to a nonlinear wave equation by a suitable choice of parameters. In addition, a comparison is made with three other methods: the Adomian decomposition method, the variational iteration method (with Green function) and the homotopy perturbation method. The matrix BLUES function method is shown to be a worthwhile alternative for the other methods.
\end{abstract}


\end{frontmatter}


\section{Introduction}
\label{sec:intro}

When studying nonlinear differential equations (DEs), it is often necessary to construct approximate analytical solutions to the equation at hand because an exact expression is not available. Hence, techniques such as the Adomian decomposition method (ADM) \cite{adomian,ADOMIAN199017}, variational iteration method (VIM) \cite{HE20073}, variational iteration method with Green function (GVIM) \cite{KHURI201428}, homotopy perturbation method (HPM) \cite{HE1999257} or homotopy analysis method (HAM) \cite{ham} have been developed and have been successfully applied to a wide range of problems in physics and other sciences.  

Recently, it has been demonstrated that the theory of Green functions can be usefully extended to inhomogeneous nonlinear ordinary differential equations (ODEs), effectively using superposition beyond the linear domain \cite{BLUES,berx,Berx_2020}. Consequently, we probed the usefulness of the BLUES function method in the arena of nonlinear partial differential equations (PDEs) with a first-order time derivative where the initial condition plays the role of the inhomogeneous source by multiplication with a Dirac delta point source located at time $t=0$ \cite{Berx2021_PDE}. It was shown that the method produces globally convergent approximants with high accuracy for systems of coupled nonlinear first-order ODEs when the fixed points of the system are introduced into the linear operator \citep{Berx2021_SIRS}. 
However, for second-order time derivatives, the BLUES function method needs to be modified. In this work we will show that the combination of the BLUES function method for PDEs with a matrix formalism is required to generate approximate solutions to PDEs with second- (or higher-) order time derivatives.

The setup of this paper is as follows. In Section \ref{sec:BLUES} we extend the BLUES function method to $n$th-order time derivatives and introduce the required matrix formalism. Next, in Section \ref{sec:telegrapher}, we apply the matrix BLUES function method to a nonlinear telegrapher equation and compare the approximants with those of the VIM, GVIM, ADM and HPM. Finally, in Section \ref{sec:conclusions} we present some conclusions and thoughts on future research.

\section{BLUES function method for $n$th-order time derivatives}
\label{sec:BLUES}
For nonlinear ODEs with source or sink terms, the BLUES function method has been studied intensively \citep{BLUES, berx, Berx_2020,Berx2021_SIRS} and anticipated in \cite{smets}.
For PDEs with a first-order time derivative, the role of the source was taken up by the initial condition multiplied by a Dirac point source located at $t=0$ \citep{Berx2021_PDE}. For $n$th-order time derivatives this approach can be extended to include the initial conditions of the derivatives of the solution. We propose the following approach: the $n$th-order in time nonlinear PDE 
\begin{equation}
\label{eq:nonlinear_PDE_scalar}
    \nloptilde u = \chi(x,t)
\end{equation}
with source $\chi(x,t)$ can be decomposed into $n$ first-order coupled PDEs where the initial conditions for the solution and the derivatives can be included as sources in their respective constituent equations by a suitable multiplication with a point source at $t=0$. The BLUES method can subsequently be applied to the system of first-order in time equations.

For equation \eqref{eq:nonlinear_PDE_scalar}, this system can be written as a different nonlinear operator $\nlop$ acting on a vector of solutions $\vect{U} = (U_1,U_2,...,U_n)$, i.e.,
\begin{equation}
    \label{eq:nonlinear_operator_psi}
    \nlop \vect{U}(x,t) = \vect{\psi}(x,t),\; \forall t>0
\end{equation}
wherein $U_1 = u$, $U_2 = \partial u/\partial t$, ..., $U_n = \partial^n u/\partial t^n$. The initial conditions for $t=0$ are collected in the vector $\vect{C}$,
\begin{equation}
    \label{eq:boundary_conditions}
    \vect{U}(x,0) = \vect{C}(x)\,.
\end{equation}
Now, $\vect{\psi}(x,t) = (0,...,0,\chi(x,t))$ is a vector of length $n$ that contains the external source in the last entry and zeroes everywhere else.
We now judiciously decompose the nonlinear operator $\nlop$ into a linear operator $\lop$, which contains time derivatives of at most first order as a consequence of the decomposition, and a residual operator $\rop$, i.e., $\rop \equiv \lop - \nlop$, which contains at least the nonlinear part of $\nlop$ and which has no explicit time dependence. Thus the action of the linear operator on $\vect{U}$ results in the following associated linear coupled system
\begin{equation}
    \label{eq:linear_operator}
    \lop\vect{U} =  \pdv{\vect{U}}{t} - A \vect{U} = \vect{\psi},\; \forall t>0\,,
\end{equation}
where $A$ is an $n\cross n$ matrix. We now rewrite the system of DEs in an equivalent form by incorporating the initial condition through multiplication of $\vect{C}$ with a Dirac delta source $\delta(t)$ and including this term on the right-hand-side of the inhomogeneous system, i.e., 
\begin{equation}
    \label{eq:linear_operator_dirac}
    \lop\vect{U}(x,t) =  \frac{\partial \vect{U}}{\partial t}(x,t) - A \vect{U}(x,t) = \vect{\psi}(x,t)\Theta(t)+\vect{C}(x) \delta(t)\equiv \vect{\varphi}(x,t), \;\forall t\geq0\,,
\end{equation}
where we have combined the external source $\vect{\psi}\Theta$ and the ``initial condition source" $\vect{C}\delta$ into the combined source $\vect{\varphi}$.

The solution of this linear system \eqref{eq:linear_operator_dirac} is the following two-variable convolution
\begin{equation}
    \label{eq:linear_solution}
    \begin{split}
        \vect{U}(x,t) &= (G\ast\vect{\varphi})(x,t)\\
        &=\int_{0^-}^t \int_\mathbb{R} G(x-x',t-t')\left[\vect{\psi}(x',t')\Theta(t') + \vect{C}(x')\delta(t')\right]\mathrm{d}x'\mathrm{d}t',\;\;\forall t \geq 0
    \end{split}
\end{equation}
where $G(x,t)$ is the Green function matrix for the inhomogeneous linear system. This $G(x,t)$ satisfies 
\begin{equation}
    \label{eq:linear_operator_delta_G}
    \lop G(x,t) = 0, \; \mbox{for} \; t >0,
\end{equation}
with Dirac-delta initial condition
\begin{equation}
    \label{eq:linear_operator_delta_initial_G}
     {\rm lim}_{t\rightarrow 0} \, G(x,t) = \delta (x)\mathbb{1}\,.
\end{equation}

Adopting the BLUES function strategy, a solution to the nonlinear system \eqref{eq:nonlinear_operator_psi} is now proposed in the form of a convolution $\vect{U}(x,t) = (B\ast\vect{\phi})(x,t)$, in which the function $B(x,t)$, named BLUES function, is taken to be equal to the Green function of the chosen related linear system, i.e., $B(x,t) \equiv G(x,t)$ and the new (combined) source $\vect{\phi}(x,t)$ is to  be calculated by systematic iteration, using the given (combined) source $\vect{\varphi}(x,t)$. This procedure starts from the following implicit equation, which makes use of the action of the residual operator,
\begin{equation}
    \label{eq:residual_operator_action}
    \begin{split}
        \rop (B\ast\vect{\phi}) &= \lop(B\ast\vect{\phi}) - \nlop(B\ast\vect{\phi})\\
        &= \vect{\phi} - \vect{\varphi}\,.
    \end{split}
\end{equation}
To find the solution to the nonlinear system \eqref{eq:nonlinear_operator_psi}, equation \eqref{eq:residual_operator_action} can be iterated to calculate an approximation for $\vect{\phi}$ in the form of a sequence in powers of the residual $\rop$. By subsequently taking the convolution product with $B(x,t)$, approximate solutions $\vect{U}_\phi^{(n)}(x,t)$ to \eqref{eq:nonlinear_operator_psi} can be found, i.e., 
\begin{equation}
    \label{eq:nth_order}
    \vect{U}^{(n)}_\phi(x,t) = (B\ast\vect{\phi}^{(n)})(x,t)= \vect{U}^{(0)}_\phi(x,t) + \left(B\ast\rop  \vect{U}^{(n-1)}_\phi\right)(x,t)\, ,
\end{equation}
where
\begin{equation}
    \label{eq:zeroth_order}
    \vect{U}^{(0)}_\phi(x,t) = (B\ast\vect{\phi}^{(0)})(x,t)= (B\ast\vect{\varphi})(x,t).
\end{equation}
is the zeroth-order convolution product in which $\vect{\phi}^{(0)}\equiv\vect{\varphi}$. We now apply this to a second-order-in-time PDE.

\section{Nonlinear telegrapher equation}\label{sec:telegrapher}
From now on we will work with dimensionless functions and variables (which are obtained after suitable scaling). The linear telegrapher equation with $\alpha,c\in\mathbb{R}$
\begin{equation}
    \label{eq:telegrapher_linear}
    u_{tt}(x,t)+\alpha u_t(x,t) -c^2 u_{xx}(x,t)= 0\,,
\end{equation}
with $u_{tt}\equiv \partial^2 u/\partial t^2$, $u_{t}\equiv \partial u/\partial t$, etc.,  is used in a broad spectrum of scientific disciplines, ranging from electrical transmission in cables, where $u$ is then the voltage per unit length of cable, to the statistical mechanics of active matter \cite{demaerel2018,WEISS2002} and mathematical biology \cite{alharbi2018}, where $u$ represents a particle or population density, respectively. Equation \eqref{eq:telegrapher_linear} interpolates between the wave equation ($\alpha \rightarrow 0$ with $c$ fixed) and the diffusion equation ($\alpha\rightarrow\infty$ and $c\rightarrow\infty$, with $c^2/\alpha\rightarrow D$ constant) and possesses a finite propagation speed for disturbances, in contrast to the infinite propagation speed for the diffusion equation \cite{evans10}, which is a result of the solution being strictly positive on $\mathbb{R}$ for all $t>0$. If at $t=0$ one assumes e.g., that $u(x,0)=\delta(x)$ then immediately after $t=0$, the solution jumps to a nonzero positive value over the entire real line, which is not compact. The information contained in the initial condition is instantaneously transported over the entire real line. Consequently, the propagation speed is infinite for the diffusion equation.

For early times $t\ll T$ where $T=1/\alpha$, the solution of equation \eqref{eq:telegrapher_linear} is wave-like, while for later times $t\gg T$, the solution is diffusive.

We now consider the telegrapher equation with a quadratic nonlinearity and constant forcing, i.e.,
\begin{equation}
\label{eq:telegrapher}
    \begin{split}
        \nloptilde u &= u_{tt}+\alpha u_t -c^2 u_{xx}+u^2 = 1\,,\\
        u(x,0) &= f(x)\,,\\
        \left.u_t(x,t)\right|_{t=0} &= g(x)\,,
    \end{split}
\end{equation}
where $f(x)$ and $g(x)$ are the initial conditions and $\chi(x,t)=1$ is the external source. We will now attempt to generate approximate solutions to \eqref{eq:telegrapher} using the formalism developed in section \ref{sec:BLUES}.

We can convert this second-order-in-time PDE \eqref{eq:telegrapher} into two coupled first-order-in-time PDEs by introducing $v = u_t$, i.e., 
\begin{equation}
\label{eq:telegrapher_matrix}
    \begin{split}
        u_t &= v\\
        v_t &= c^2 u_{xx} - \alpha v - u^2 +1\,,
    \end{split}
\end{equation}
where $\vect{\psi}(x,t) = (0,1)^\intercal$. We can rewrite the system \eqref{eq:telegrapher_matrix} with solution vector $\vect{U} = (u,v)^\intercal$ in the form required by the matrix BLUES function method \eqref{eq:nonlinear_operator_psi},
\begin{equation}
    \label{eq:telegrapher_vectorial}
    \nlop\vect{U}(x,t) = \vect{\varphi}(x,t)\,.
\end{equation}
The linear operator $\lopt$ (which is now independent of $x$), residual $\rop$ and source vector $\vect{\varphi}$ are given through the following expressions
\begin{equation}
    \label{eq:operators_and_source}
    \begin{split}
    \lopt\vect{U} &= \begin{pmatrix}
        u_t \\
        v_t
    \end{pmatrix} - A \begin{pmatrix}
        u \\
        v
    \end{pmatrix},\\
    \rop\vect{U} &= \begin{pmatrix}
        0 \\
        c^2 u_{xx} -u^2
    \end{pmatrix},\\
    \vect{\varphi} &= \delta(t)\begin{pmatrix}
        f(x) \\
        g(x)
    \end{pmatrix} + \begin{pmatrix}
    0 \\
    1
    \end{pmatrix}\,.\\
    \end{split}
\end{equation}
The matrix of coefficients $A$ of the linear operator is
\begin{equation}
    \label{eq:linear_coefficients}
    A = \begin{pmatrix}
        0 & 1\\
        0 & -\alpha
    \end{pmatrix}\,,
\end{equation}
which is independent of $x$, and hence the matrix Green function for the linear operator is the matrix exponential $G(x,t) = \delta(x)\exp(At)\Theta(t)$, i.e.,
\begin{equation}
    \label{eq:green_matrix}
    G(t) = \delta(x)\begin{pmatrix}
        1 & \frac{1-\me^{-\alpha t}}{\alpha} \\
        0 & \me^{-\alpha t}
    \end{pmatrix}\Theta(t)\,.
\end{equation}

We can now set up the BLUES iteration procedure. The $n$th approximant for the solution of \eqref{eq:telegrapher_matrix} is calculated by
\begin{equation}
    \label{eq:iteration_telegrapher}
    \vect{U}^{(n)}(x,t) = \vect{U}^{(0)}(x,t) + \int_{0^-}^t \int_\mathbb{R} G(x-x',t-t')\ropx \vect{U}^{(n-1)}(x',t')\mathrm{d}x'\mathrm{d}t'\,,
\end{equation}
where $\vect{U}^{(0)}(x,t)$ is the zeroth approximant and is defined as the convolution product of the Green function and the source $\vect{\varphi}(x,t)$, i.e, 
\begin{equation}
    \label{eq:zeroth-order}
    \vect{U}^{(0)}(x,t) = (G\ast\vect{\varphi})(x,t) = \left(f(x)+\frac{t}{\alpha}+\frac{1-\alpha g(x)}{\alpha^2}(\me^{-\alpha t} -1), \frac{1}{\alpha} - \frac{1-\alpha g(x)}{\alpha}\me^{-\alpha t}\right)^\intercal\,.
\end{equation}

For example, for initial conditions $f(x) = 1+\sin x$ and $g(x) =0$, the zeroth approximant is (for $\alpha=1$)
\begin{equation}
    \label{eq:zeroth-order_filled}
    \vect{U}^{(0)}(x,t) = \left(\sin x + t+\me^{-t}, 1-\me^{-t}\right)^\intercal
\end{equation}
and the first approximant $\vect{U}^{(1)}(x,t) = (u^{(1)}(x,t),v^{(1)}(x,t))$ consists of
\begin{equation}
    \label{eq:first-order_filled}
    \begin{split}
        u^{(1)}(x,t) &= \sin{x}\left[\me^{-t} (2 t - (c^2 -4)) -t^2 - (c^2-2)t + (c^2-3)\right]+ \frac{1}{2}\cos{2x}\left[\me^{-t} + t-1\right]\\
        &-\frac{1}{6}\left[3\me^{-2t} - \me^{-t} (6t^2 + 12t+9) + (t^3 - 6t^2 +9t)\right]\,,\\
        v^{(1)}(x,t) &= u^{(1)}_t(x,t)\,.
    \end{split}
\end{equation}
Note that when a higher-order-in-time PDE is converted into a system of coupled PDEs, the residual is zero in the individual channels of the ``new'' variables, i.e., the derivatives. The matrix formalism therefore decouples into individual integrations of the elements in the last column of the Green function matrix with the residual applied to the previous approximant for the solution $u(x,t)$. Hence, it is only necessary to perform the integration corresponding to the channel for the solution $u(x,t)$ and not for all the time derivatives. However, this is not the case when the system is \emph{a priori} coupled.

\begin{figure}[htp]
    \centering
    \includegraphics[width=0.85\linewidth]{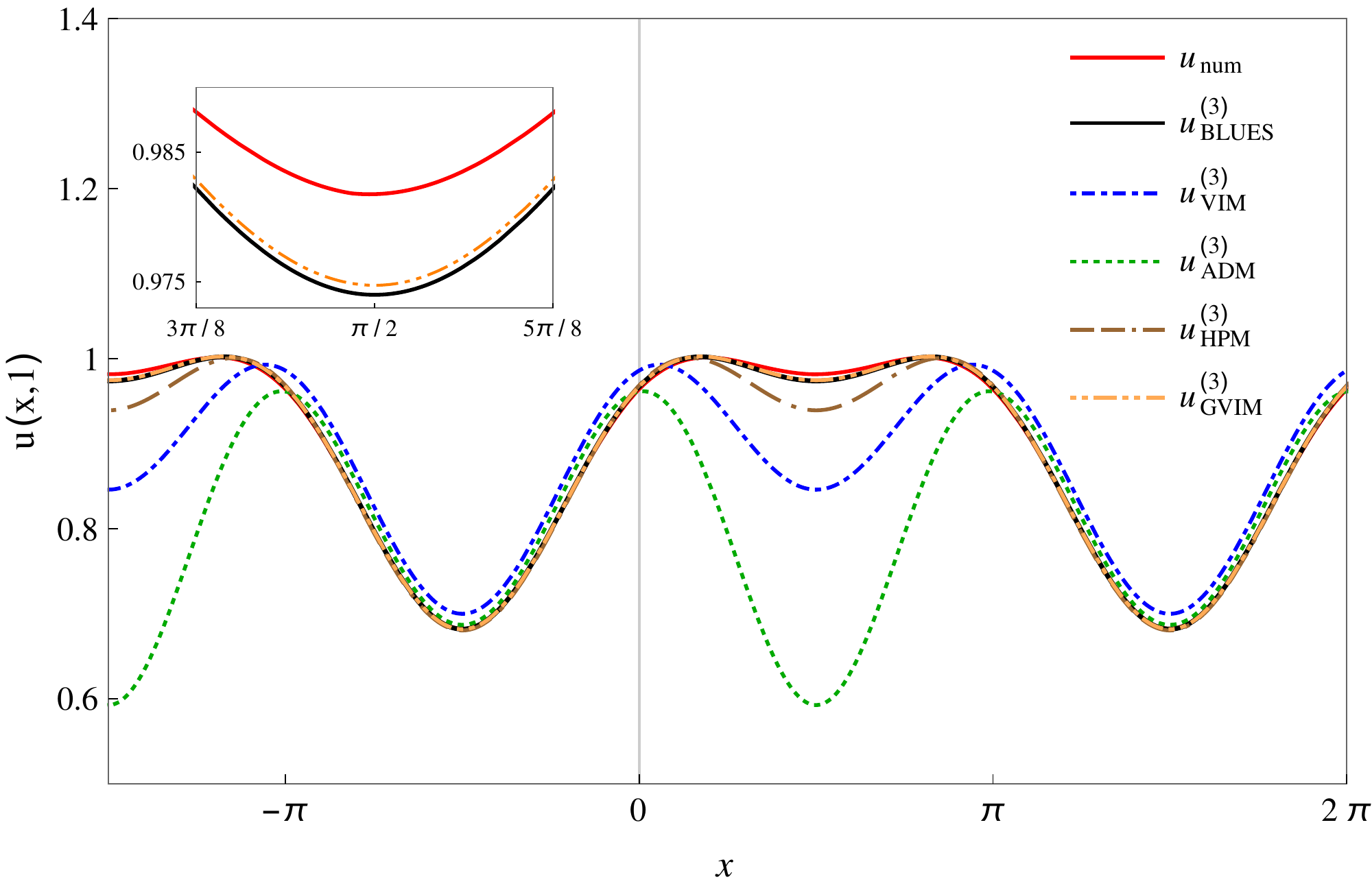}
    \caption{A comparison is made between the numerical solution of equation \eqref{eq:telegrapher} (red, full line), the matrix BLUES function method (black, full line), the VIM (blue, dot-dashed line), the ADM (green, dotted line), the HPM (brown, dot-dash-dashed line) and the GVIM (orange, dot-dot-dashed line). For all methods the third iteration or third order result is shown. The inset shows details around the local minimum at $x=\pi/2$. The time is fixed at $t=1$ and parameter values are $\alpha = c =1$.}
    \label{fig:telegrapher_X}
\end{figure}

\begin{figure}[htp]
    \centering
    \includegraphics[width=0.85\linewidth]{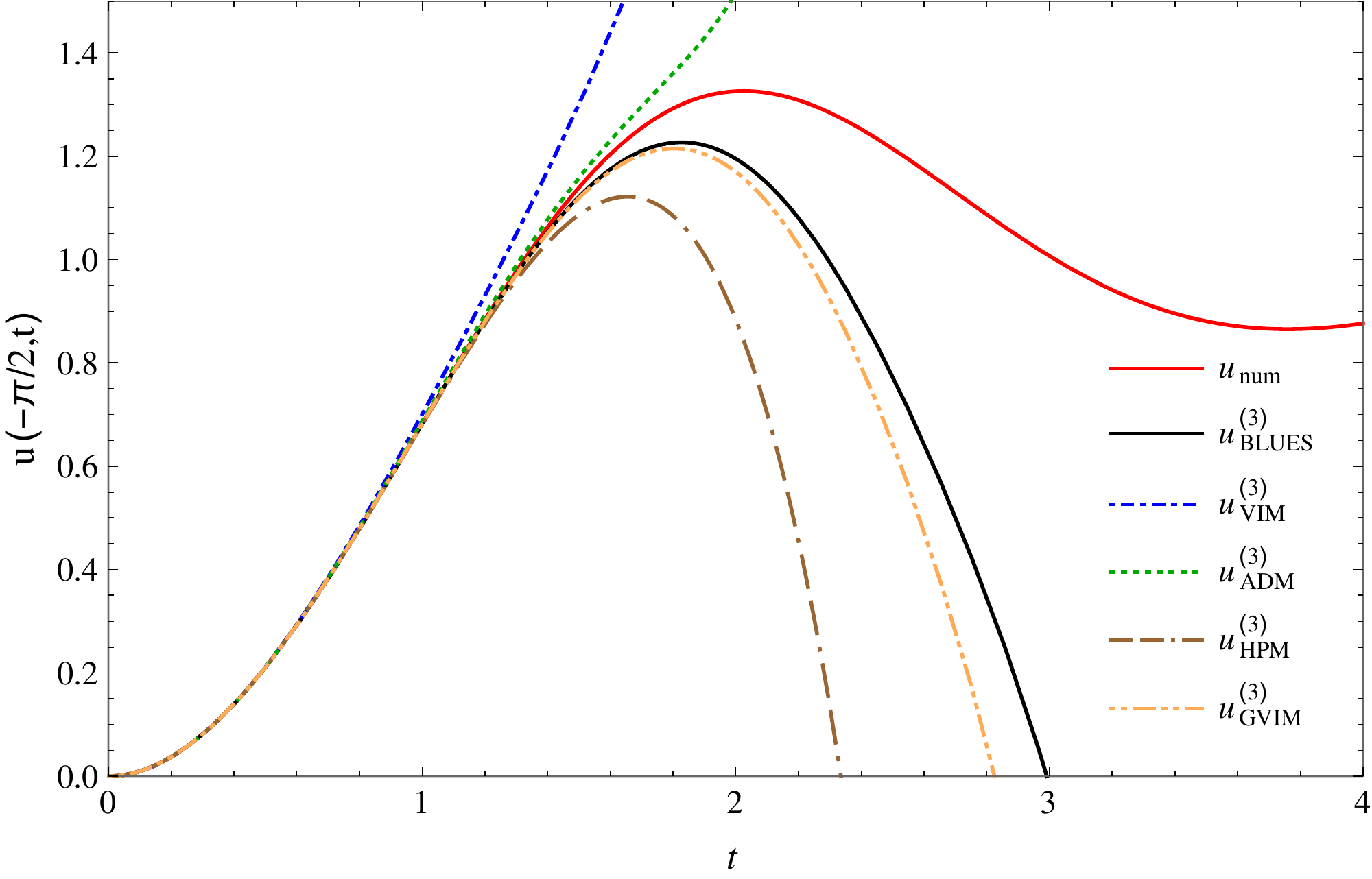}
    \caption{A comparison is made between the numerical solution of equation \eqref{eq:telegrapher} (red, full line), the matrix BLUES function method (black, full line), the VIM (blue, dot-dashed line), the ADM (green, dotted line), the HPM (brown, dot-dash-dashed line) and the GVIM (orange, dot-dot-dashed line). For all methods the third iteration or third order result is shown. The position is fixed at $x=-\pi/2$ and parameter values are $\alpha = c =1$.}
    \label{fig:telegrapher_T}
\end{figure}

In Figs. \ref{fig:telegrapher_X} and \ref{fig:telegrapher_T}, we compare the approximate solutions of the matrix BLUES function method with the VIM, ADM, HPM and GVIM for $\alpha = 1$ and $c=1$, with initial conditions $f(x) = 1+\sin x$ and $g(x) =0$. All numerical results are obtained with Mathematica's NDSolve function, which uses the numerical method of lines to discretize the PDE in all but one dimension and then integrates this semi-discrete problem as a system of ODEs. In Fig. \ref{fig:telegrapher_X}, the spatial profile of the approximants is shown for a fixed time $t=1$. The matrix BLUES function method and the GVIM, which are almost concident, approximate the numerical solution quite well. This is confirmed in Fig. \ref{fig:telegrapher_T}, where the time evolution of the approximants for fixed position $x = -\pi/2$ is shown. Note, however, that all of the implemented methods diverge for $t\rightarrow\infty$. The accuracy can be increased by considering higher-order iterations. Note that while all of the methods can quite accurately reproduce the spatial profile at the \textsl{global} minima of the numerical solution, the performance differs significantly in the region where the maxima and \textsl{local} minima occur (e.g., around $x=\pi/2$).  

\begin{figure}
    \centering
    \includegraphics[width=0.85\linewidth]{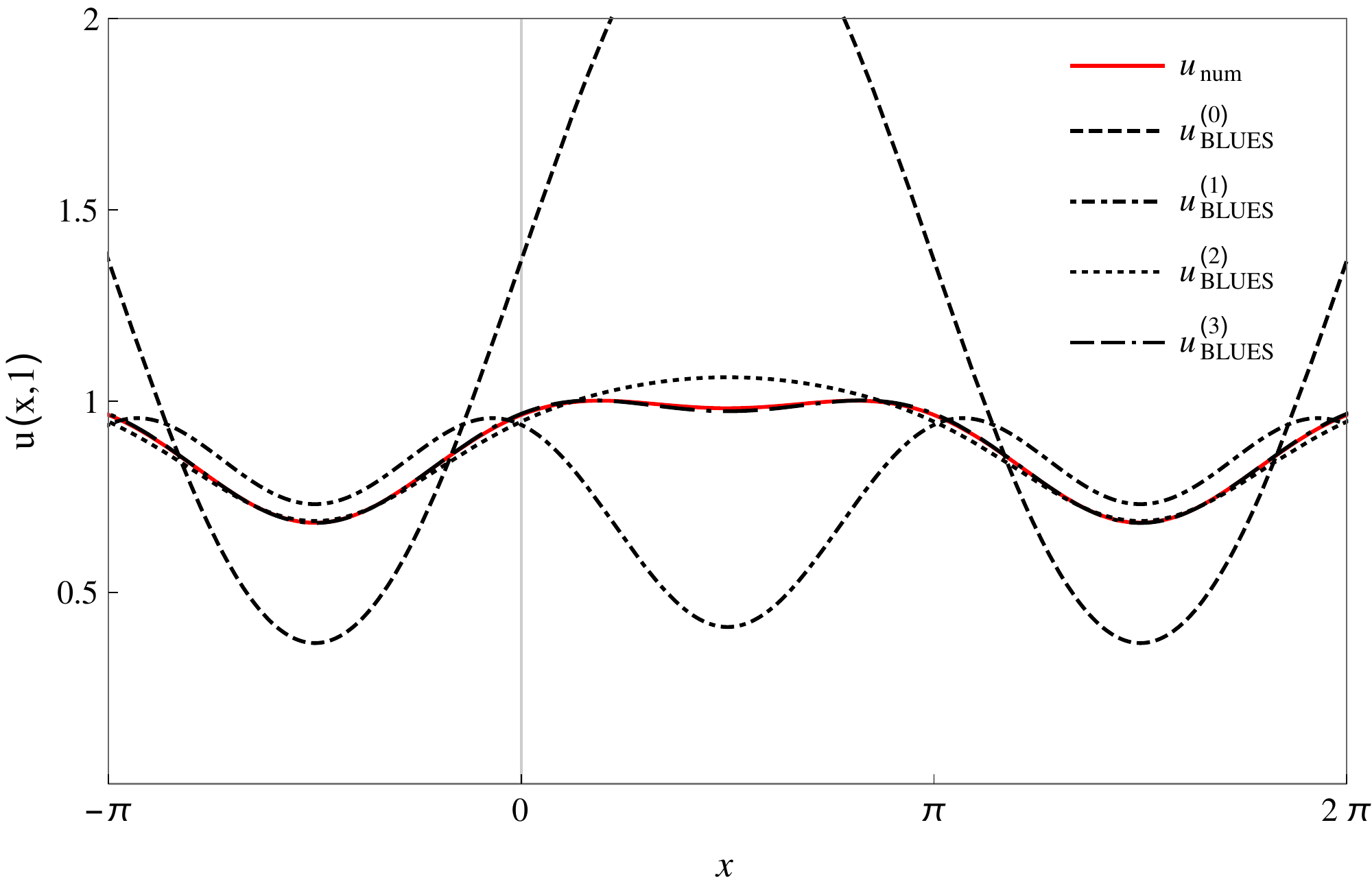}
    \caption{Spatial plot of the solution of the nonlinear telegrapher equation \eqref{eq:telegrapher} for a fixed time $t=1$. A comparison is made between the numerical solution (red, full line) and the  $n\in\{0,1,2,3\}$ matrix BLUES method approximants (black lines). The parameter values are $\alpha = c =1$.}
    \label{fig:blues_approximants}
\end{figure}
In Fig. \ref{fig:blues_approximants}, the consecutive matrix BLUES method approximants for $n\in\{0,1,2,3\}$ are compared with the numerical solution at $t=1$. Note that the local minimum around $x=\pi/2$ as well as the entire solution is well reproduced by the third approximant.

\section{Conclusions}
\label{sec:conclusions}
In this paper we have extended the BLUES function method that was developed for nonlinear PDEs with a first-order time derivative to second-order (and higher) time derivatives by introducing a matrix formalism. We have applied this extension of the BLUES function method to a second-order-in-time nonlinear telegrapher equation with a quadratic nonlinearity and a trigonometric initial condition. From a comparison with the ADM, the VIM, the HPM and the GVIM, we have shown that the matrix BLUES function method is able to generate approximants that are a valuable alternative to what can be obtained through other methods at the same level of iteration or the same order of approximation. In future research on the BLUES function method, we envision a possible extension of the method to stochastic (coupled) DEs for which the noise can play the role of an external source. 

\bibliographystyle{elsarticle-num}
\bibliography{BLUES.bib}

\end{document}